\documentclass[sigconf]{acmart}
\usepackage{subcaption}

\AtBeginDocument{%
  }

\setcopyright{acmlicensed}
\copyrightyear{2018}
\acmYear{2018}
\acmDOI{XXXXXXX.XXXXXXX}
\acmConference[Conference acronym 'XX]{Make sure to enter the correct
  conference title from your rights confirmation email}{June 03--05,
  2018}{Woodstock, NY}
\acmISBN{978-1-4503-XXXX-X/2018/06}

\usepackage{float}

\newcommand{\out}[1]{}

\begin{document}

\title{SARCH: Multimodal Search for Archaeological Archives}


\author{Nivedita Sinha}
\affiliation{
    \institution{IIT Delhi, New Delhi}
   \country{India}
}
\author{Bharati Khanijo}
\affiliation{
    \institution{IIT Delhi, New Delhi}
   \country{India}
}
\author{Sanskar Singh}
\affiliation{
    \institution{IIT Delhi, New Delhi}
   \country{India}
}
\author{Priyansh Mahant}
\authornote{Work done while interning at IIT Delhi}
\affiliation{
    \institution{CSV Technical University, Bhilai}
   \country{India}
}
\author{Ashutosh Roy}
\authornotemark[1]
\affiliation{
    \institution{CSV Technical University, Bhilai}
   \country{India}
}

\author{Saubhagya Singh Bhadouria}
\authornotemark[1]
\affiliation{
    \institution{GGSIP University, New Delhi}
   \country{India}
}
\author{Arpan Jain}
\authornotemark[1]
\affiliation{
    \institution{AK Technical University, Lucknow}
   \country{India}
}
\author{Maya Ramanath}
\affiliation{
    \institution{IIT Delhi, New Delhi}
   \country{India}
}

\renewcommand{\shortauthors}{Sinha et. al.}

\begin{abstract}
In this paper, we describe a multi-modal search system designed to search old archaeological books and reports. This corpus is digitally available as scanned PDFs, but varies widely in the quality of scans. Our pipeline, designed for multi-modal archaeological documents, extracts and indexes text, images (classified into maps, photos, layouts, and others), and tables. We evaluated different retrieval strategies, including keyword-based search, embedding-based models, and a hybrid approach that selects optimal results from both modalities. We report and analyze our preliminary results and discuss future work in this exciting vertical.
\end{abstract}

\maketitle

\section{Introduction}
\begin{figure}[h]
\begin{subfigure}[b]{0.4\columnwidth}
    \centering
    \includegraphics[width=\linewidth]{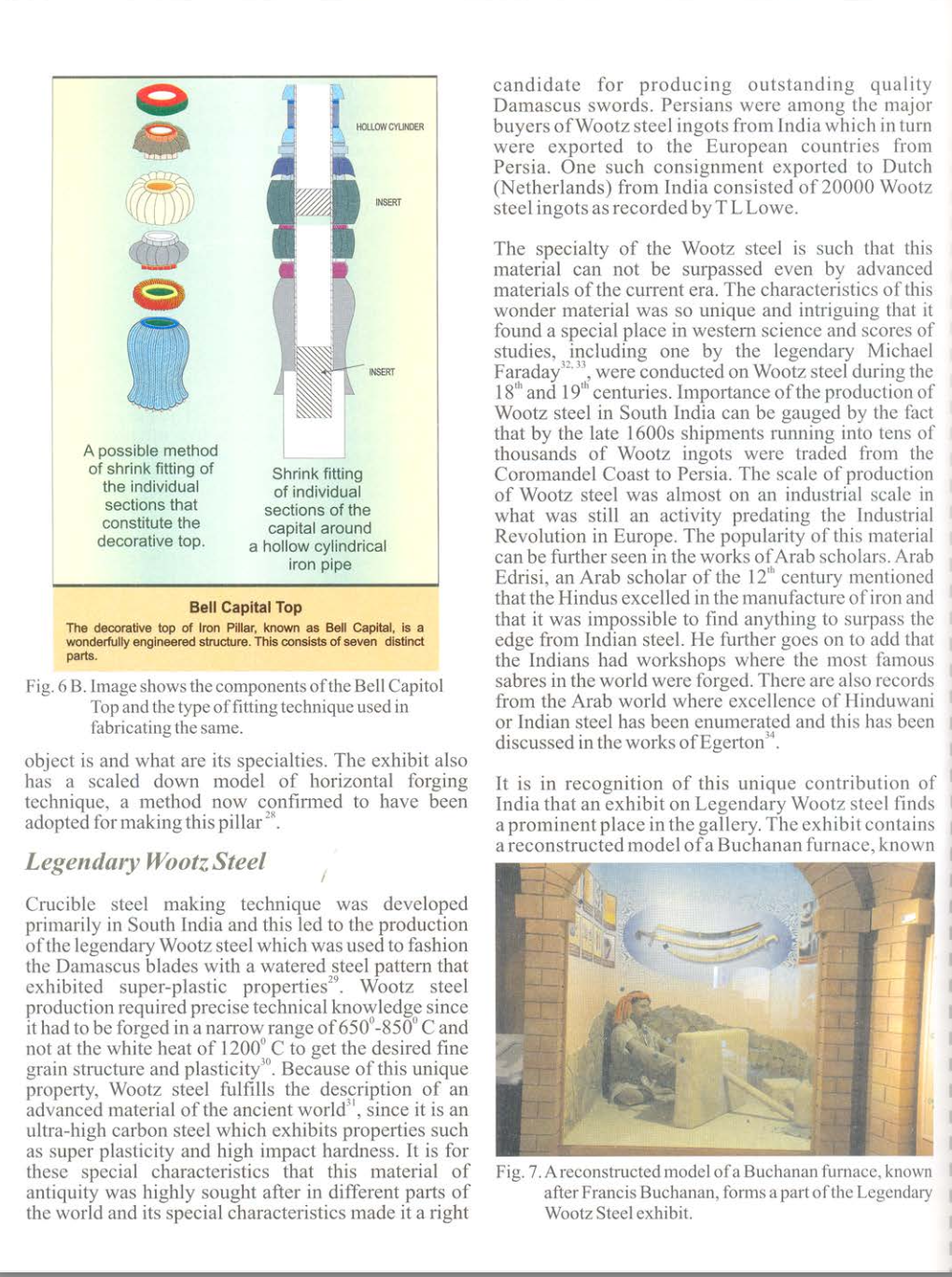}
    \caption{Image showing 2-column text, image with caption}
    \label{fig:scanned-page}
\end{subfigure}
\hfill
\begin{subfigure}[b]{0.4\columnwidth}
    \centering
    \includegraphics[width=\linewidth]{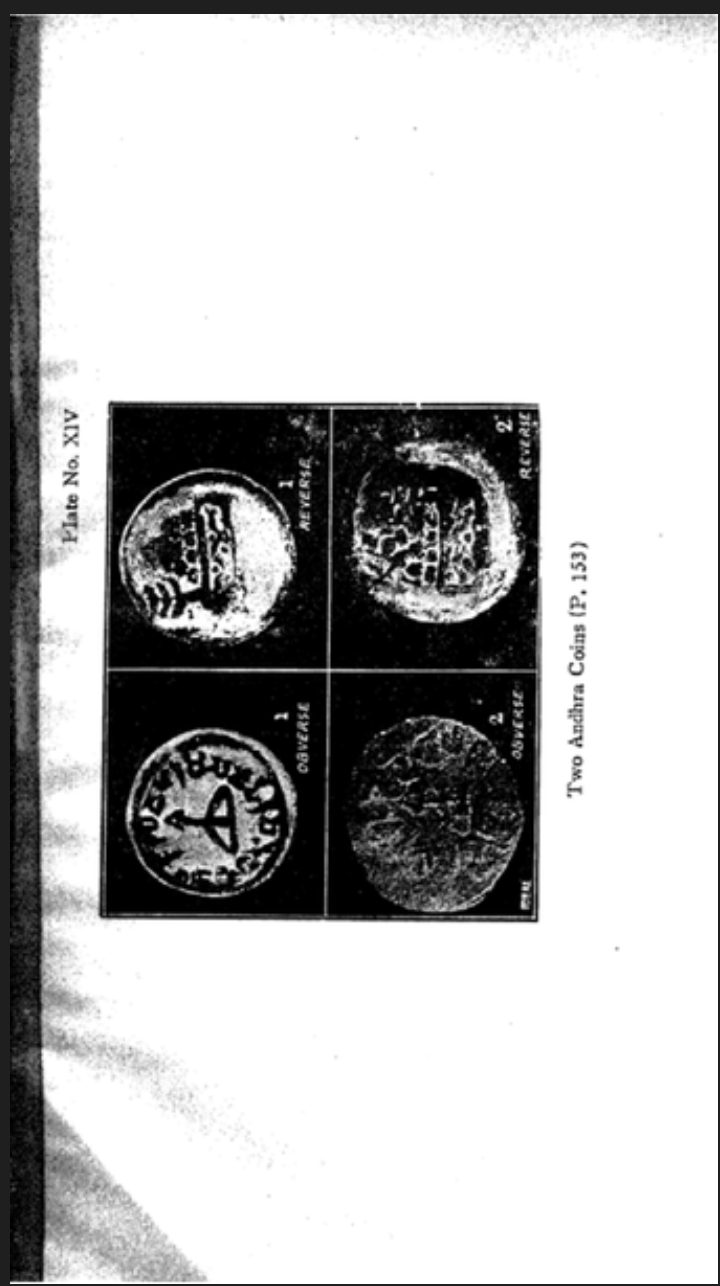}
    \caption{Page with shadows and rotated image}
    \label{fig:scanned-page2}
    
\end{subfigure}
\caption{Complex layouts and bad quality scans}
\label{fig:scans}
\end{figure}

Old archaeological archives, often consisting of field reports, survey documents, excavation records and books are invaluable resources for researchers. These documents are typically stored as scanned pdfs, consisting of multi-modal data, including textual descriptions, maps, photos and tables, that provide rich historical and geographical context. While many corpora have multi-modal content,  there are a number of \emph{archaeology-specific} challenges when dealing with old archaeological records.
\begin{itemize}
    \item \emph{Complex layouts}: Figure \ref{fig:scanned-page} shows a rather complicated layout with 2-column text with 2 figures, of which one Figure (left/top) has text of its own that describes the images and should be extracted. Further, Figure \ref{fig:scanned-page2} shows a bad quality scan that prevents users from observing the details of the photos of coins. The photo itself is oriented differently then the rest of the document.

    \item \emph{Multiple modalities}: While prior work has typically dealt with models for text, images and tables, we find that this categorisation is insufficient when dealing with archaeological data. First, "image" is too broad a category -- we need to further break it down into specific kinds of images, such as \emph{maps, photos, layouts}, all of which are meaningful in the archaeological domain. Second, textual content from \emph{within} the image or table adds very important context. This can specially be seen in the case of maps -- the locations in the maps are important context and serve to describe the map itself. Third, image or table captions are often short and not very descriptive. The actual description comes from within the text that refers to the image or table. Such an example can be seen in Figure \ref{fig:scanned-page2}, where the actual description of those coins are described in a different page.

    
\end{itemize}

Table \ref{tab:examples} shows examples of text queries and results. As can be seen, the modality of the result is different for each query and is to be inferred using the query itself.


\begin{table*}[ht]
    \centering
    \begin{tabular}{|p{6cm}|c|p{4cm}|}
    \hline
       {\bf Query}  & {\bf Result type} & {\bf Relevant result in the top-5} \\
       \hline
        Map showing sites discovered during 1948 and 1957 & Map & \raisebox{-0.5\totalheight}{\includegraphics[width=\linewidth]{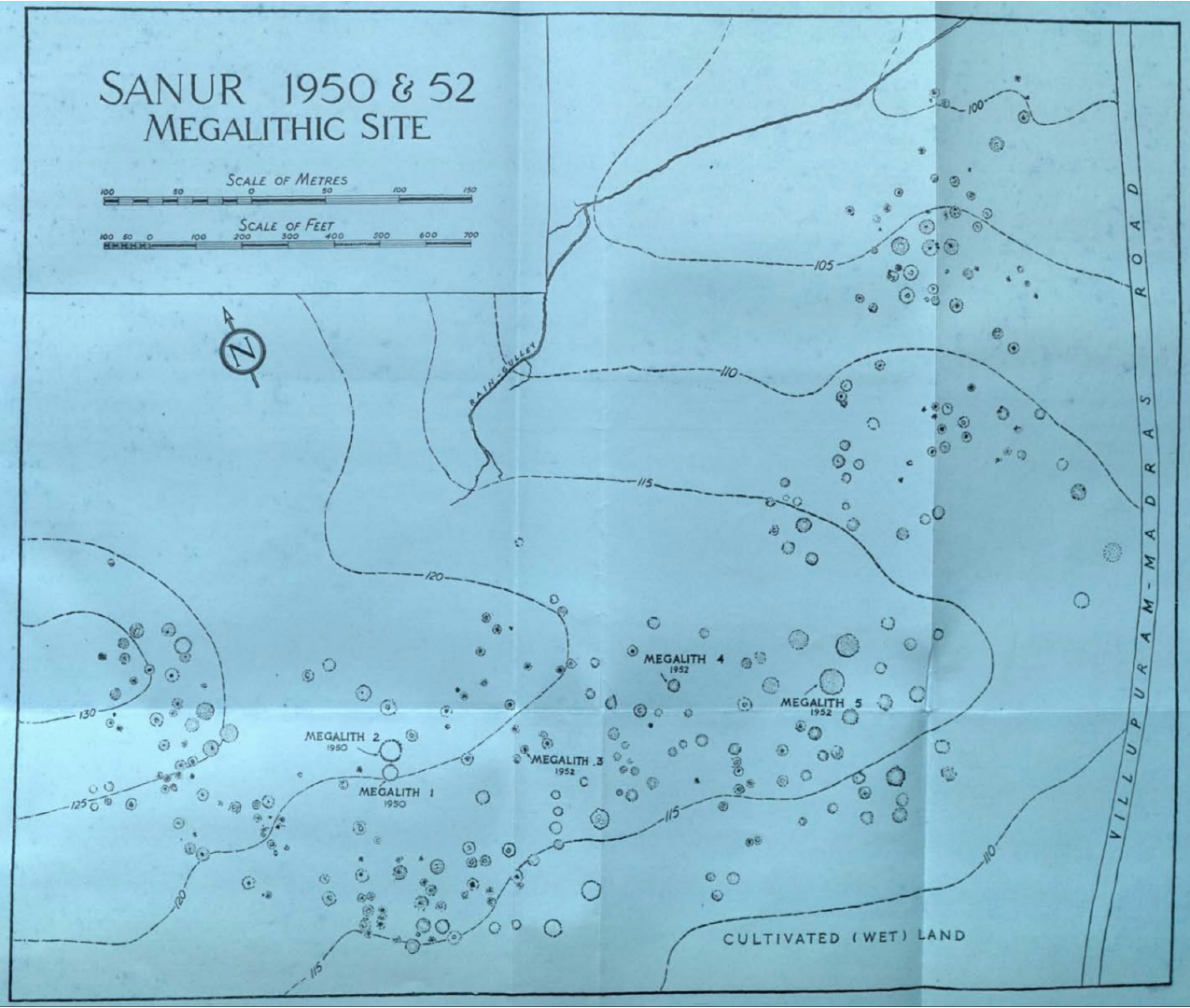}}\\
        \hline
        Primary crops of the Harappan civilization & Text & \raisebox{-0.5\totalheight}{\includegraphics[width=\linewidth]{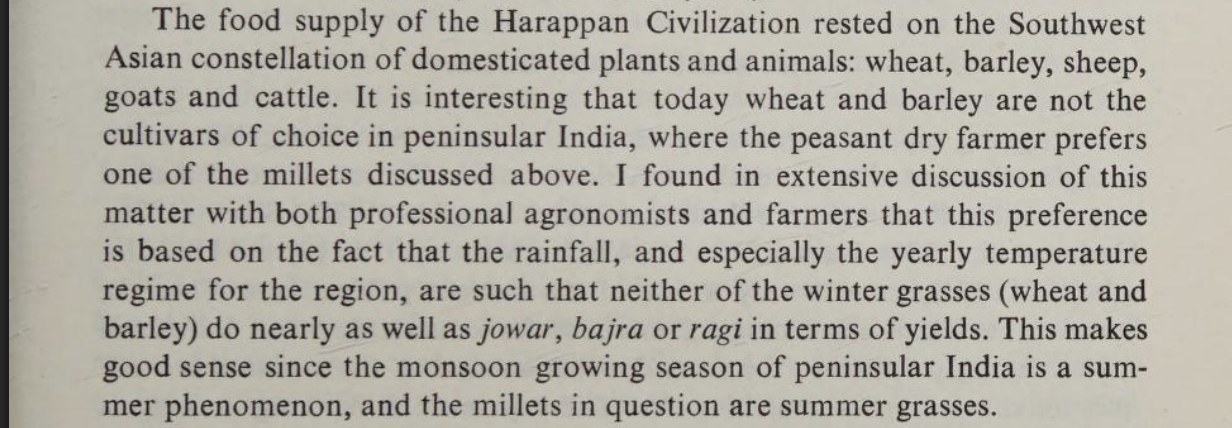}}\\
        \hline
         Raw material used for making beads?  & Table & \raisebox{-0.5\totalheight}{\includegraphics[width=\linewidth]{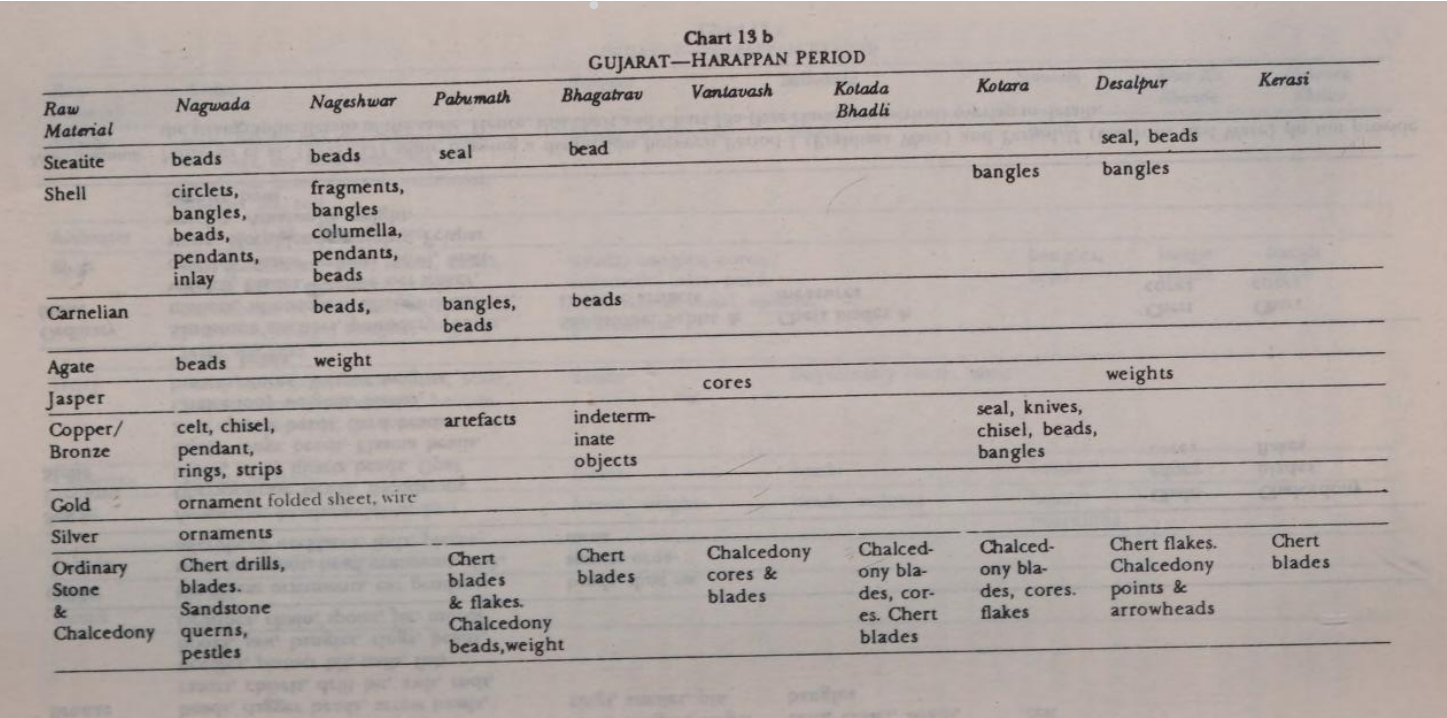}}\\
    \hline
    \end{tabular}
    \caption{Example queries in our benchmark and a top-5 result returned by our search engine}
    \label{tab:examples}
\end{table*}

In this paper, we describe our system, SARCH\footnote{Simply pronounced "search" -- Searching Archaeological Archives)}, an end-to-end solution for pre-processing archival scans of archaeological content and a search engine to search the archive in different modalities. Our pre-processing pipeline separates text from images and tables, and further classifies images into maps, layouts, photos, and a generic "figure". Further, to facilitate the search functionality, the extraction pipeline looks to add context to: i) maps, by extracting locations mentioned in the maps, and tables, by extracting the contents of the table, ii) images and tables, by associating them with their captions and additionally searching surrounding text for descriptions of the image or table. As mentioned before, the extraction is challenging because of the varying quality of scans (see Section \ref{sec:pipeline} for details). 

While typical multi-modal search may receive queries in text, but return results from other modalities, the same idea comes with additional complexity in the case of archaeological search. As mentioned earlier, low-quality scans are quite prevalent. While applying standard corrections may help in extracting text, the same may not be true of images. Therefore, most of the description of the image comes not from the image itself, but from its context (that is, the surrounding text where the image is described), thereby limiting the utility of image embeddings alone. Hence, we applies three different retrieval strategies: keyword-only, embeddings-only and a hybrrid search that does both embeddings-based search as well as keyword search (see Section \ref{sec:search} for details).  

{\bf A video demonstrating the proposed system can be accessed at https://tinyurl.com/sarchdemo.}




\subsubsection*{Related Work} 
Around the world, several extensive digital archaeological archives such as The Digital Archaeological Record \cite{tdar}, Archaeology Data Service \cite{ads}, Arachne \cite{arachne}, etc. help researchers document their research and search for related literature and artifacts within these websites. However, all these services rely extensively on manual curation and metadata for their search. In contrast, our corpus consist simply of old, scanned books and reports that are automatically processed. We rely on automated methods of text, image, and table extraction. Our search is based directly on the content of the documents, rather than metadata.

For our search system, we make use of MiniLM \cite{sentenceTransformerMiniLMPaper}, CLIP \cite{clip}, and TAPAS \cite{herzig2020tapas} for textual content, images and tables, respectively, and use the embeddings to conduct similarity searches on user queries. Further, we perform keyword search using the BM25 \cite{robertson2009probabilistic} scoring method. More details are described in Section \ref{sec:search}.

\subsection{Contributions}
In this paper, we address the problem of building a robust, multi-modal search system that enables retrieval from archival archaeological legacy documents. Our contributions are as follows:
\begin{itemize}
    \item a pipeline for extracting multi-modal content from old, scanned archaeological reports.

    \item three different search models: keyword only, embeddings only and a hybrid search system utilizing both embedded vector search as well as keyword search. A key feature of our system is that it is able to return results of types that are very specific to archaeology, such as maps, photos and layouts.

\end{itemize}

\subsubsection*{Organisation} The rest of this paper is organised as follows. Section \ref{sec:pipeline} describes the offline processing of our system, including content extraction and cleaning for scanned archaeological texts. Section \ref{sec:search} gives an overview of our search engine along with sample results in each of the three modalities, while Section \ref{sec:conc} concludes and outlines future work.

\section{Offline Processing}
\label{sec:pipeline}
\begin{figure}
    \centering
    \includegraphics[width=0.75\linewidth]{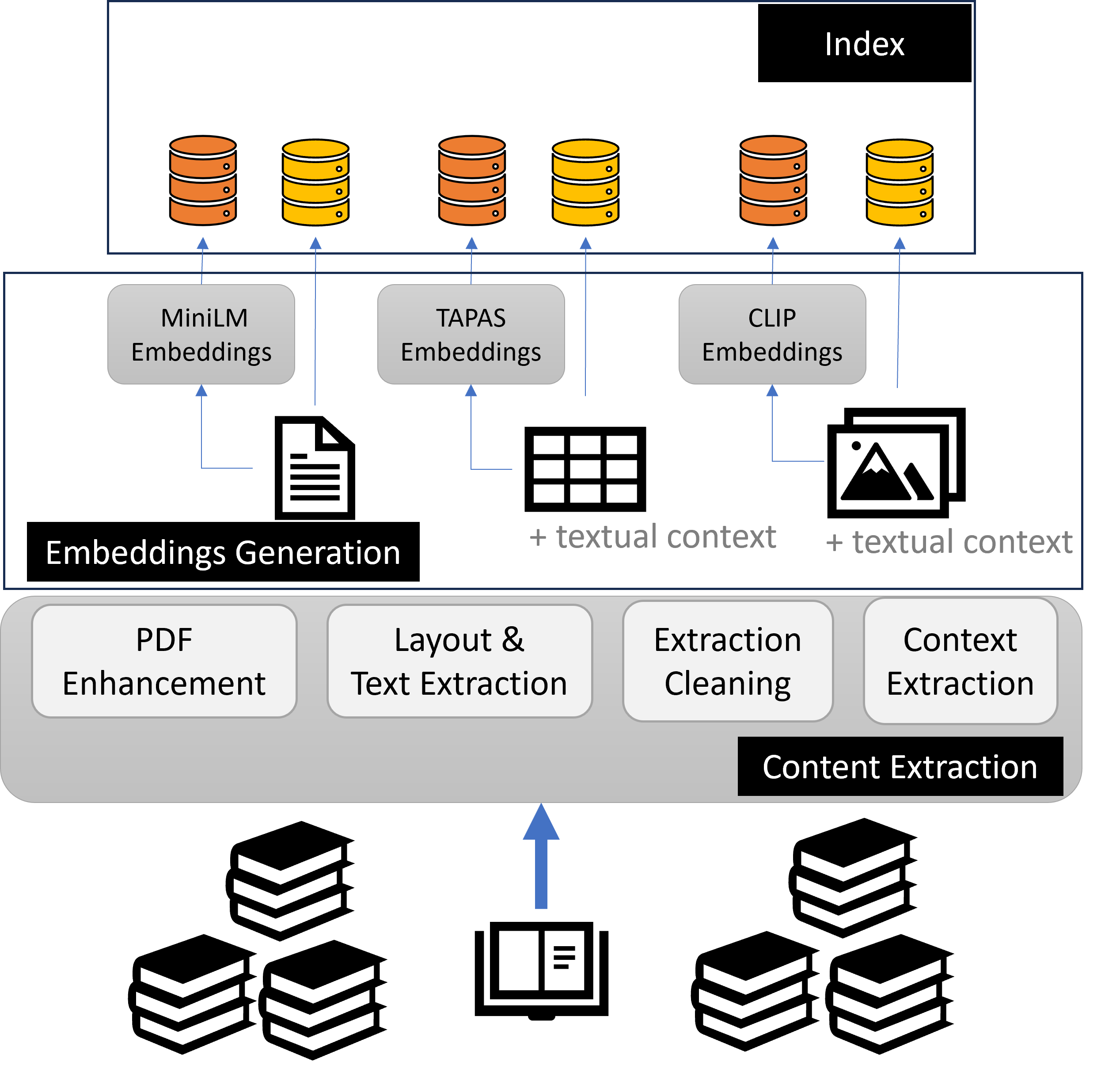}
    \caption{Offline processing: content extraction, embedding generation and indexing}
    \label{fig:ingestion}
\end{figure}

Figure \ref{fig:ingestion} shows the components of our offline processing system --  content extraction, embeddings generation and indexing. The input to the system is the corpus of scanned pdf files. Our corpus currently has 296 scanned documents, consisting of 63,208 pages. We describe these three components next.

\subsection{Content Extraction}
\subsubsection*{PDF enhancement}
The scanned pages vary greatly in quality. The use of OCR tools are best done over good quality pdfs. To enhance quality of OCR results, we processed each page as follows: i) convert the page to grayscale, ii) denoise using bilateral filtering (this is chosen for its superior edge preservation, which in turn is advantageous for scanned textual content), iii) process with Otsu’s thresholding \cite{otsu1979}. This workflow improves text clarity by producing a high-contrast binary image. All these steps are implemented with OpenCV \cite{bradski2000opencv}.

\subsubsection*{Layout and Content Extraction}
Parsing the layout of the page was a big challenge, as the layout could combine multiple modalities. We used Surya Document Analysis Toolkit \cite{suryaocr} to identify text, images, and tables. This toolkit produces bounding boxes and classifies each bounding box as text, image, or table. However, the extraction of tabular structure and content was not robust. Therefore, for table content extraction, we used Qwen2.5-VL-7B-Instruct \cite{bai2025qwen2}. Once the layout was parsed, text regions were analyzed using OCR \cite{suryaocr}. 


Additionally, we implement a CLIP \cite{clip} based zero shot classifier where an input image is classified into one of the 4 classes: map, photograph, site layout, and figure.

\subsubsection*{Extraction Cleaning}

Despite efforts to improve the scan quality, quite often the OCR tools fail in correctly extracting the text. Additionally, formatting tags like bold, italics were also present in the text extracted by OCR. These tags affect search quality since they are not part of the semantic content of the page. Therefore, we had a "text cleaning" module that took as input whatever text was extracted by the OCR tool and attempted to correct the text. We used beautiful soup python library \cite{beautifulSoup} for this purpose. Spell correction, using TextBlob \cite{textblob} was done to improve text search.

For tabular data cleaning, we first used Qwen to extract the contents of the table, but noticed that further processing was required for more accurate extraction. Therefore, we implemented our own rule-based implementation that takes as input the extracted content from Qwen and further cleans the data to ensure cell alignment with the appropriate header and to handle empty cell values. During preprocessing, HTML tags are removed from the extracted content. Empty cells in tables are filled with NaN to create a consistent format. Rows are also adjusted so they always match the number of columns in the header, ensuring a uniform table structure for later processing. 

\subsubsection*{Context Extraction} This module is the most important module for our search system. For images and tables, the context in which they appear is key to ensuring that they are returned as relevant results. The most obvious context for both images and tables is the caption. But, a more semantically rich context can be found in the text that describes the image or table, as well as text that occurs in the image or table itself. 

\begin{figure}[ht]
    \centering
    \includegraphics[width=0.5\linewidth]{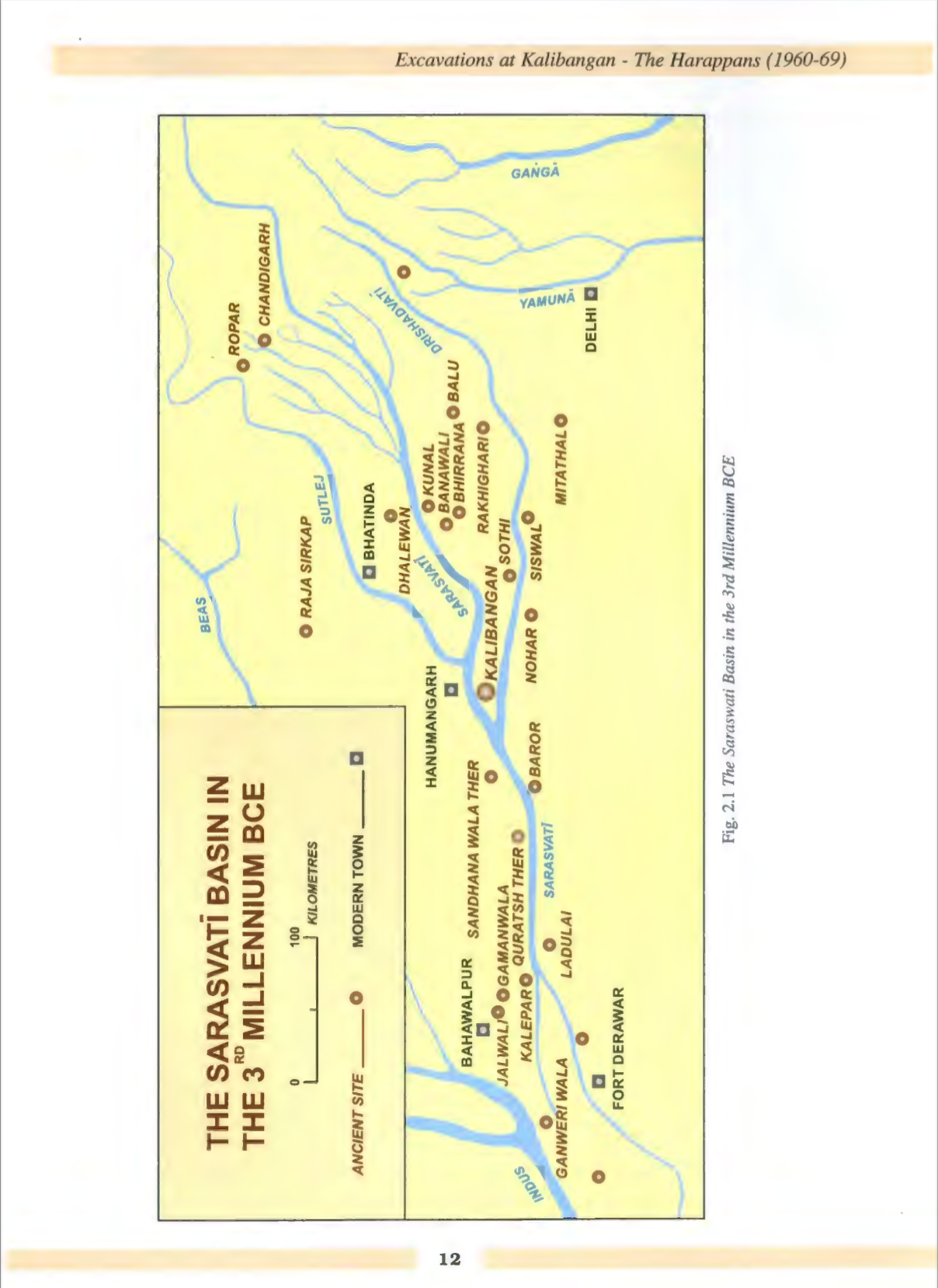}
    \caption{Rotated Picture on page}
    \label{fig:rotated}
\end{figure}

Therefore, we further enhance the context in the following ways. 
\begin{enumerate}
    \item {Sometimes a caption could be present in the text extracted within the image, especially where the image is rotated. Here, it is extracted by finding the location the word like "Fig." or "Figure" within the OCR text followed by number until the end of sentence as shown in Figure \ref{fig:rotated}.

    \item We look through the textual content surrounding the images or tables on the page where they were found, as well as the previous and next page and identify the paragraphs that mention the ordinal number of that image (table), for example, "Figure 2 shows the map of ...." through string matching. We identified the description containing the image ID, e.g. "Figure 2" present in the caption. The image may be referred to in multiple paragraphs. We use all these paragraphs as context.

    \item For {\bf maps}, we extract the location names from the map. This is a challenging task since the location names could be spelled out with non-standard spacing as well as orientations. Therefore, a simple OCR will only extract a small subset of location names. We are using Surya OCR model \cite{surya} to read any text present inside the map. Future work includes trying more sophisticated text spotting methods for maps.

    \item For {\bf tables}, the content of the tables is also used as context. Apart from row and column headers, the non-numerical values are valuable context and help in retrieval.  To process each table and produce matching captions, we use the Qwen \cite{bai2025qwen2} model. First, Qwen is prompted to extract any table-related captions or descriptions that may already exist. They are saved as the table's caption if Qwen can find them. If not, we give Qwen instructions to summarize the table's contents, which is subsequently utilized to generate a table caption. }
\end{enumerate}

\subsubsection*{Pipeline Output}
Once a pdf file goes through the pipeline, we have the following content that is ready for indexing: i) textual content along with page number, ii) images and tables, image type (relevant only for images),  associated with their contextual text and their page numbers. 

In all, we have 63,208 pages, 22,349 images, and 1,824 tables that we extracted from our corpus of 296 documents.

\subsection{Embeddings Generation and Indexing}
As mentioned in the Introduction, our search system is a hybrid system and makes use of both embeddings search as well as keyword search. As shown in Figure \ref{fig:ingestion}, for each of the three kinds of content that we extract -- text, image (further classified into map, photo, layout, and figure), and table -- separate embeddings are generated using MiniLM \cite{sentenceTransformerMiniLMPaper}, CLIP \cite{clip} and TAPAS \cite{herzig2020tapas}, respectively and stored in a database to facilitate top-$k$ nearest neighbor search. In addition, the textual content is stored as an inverted index to facilitate keyword search.

\section{Search Architecture}
\label{sec:search}
\begin{figure}
    \centering
    \includegraphics[width=0.75\linewidth]{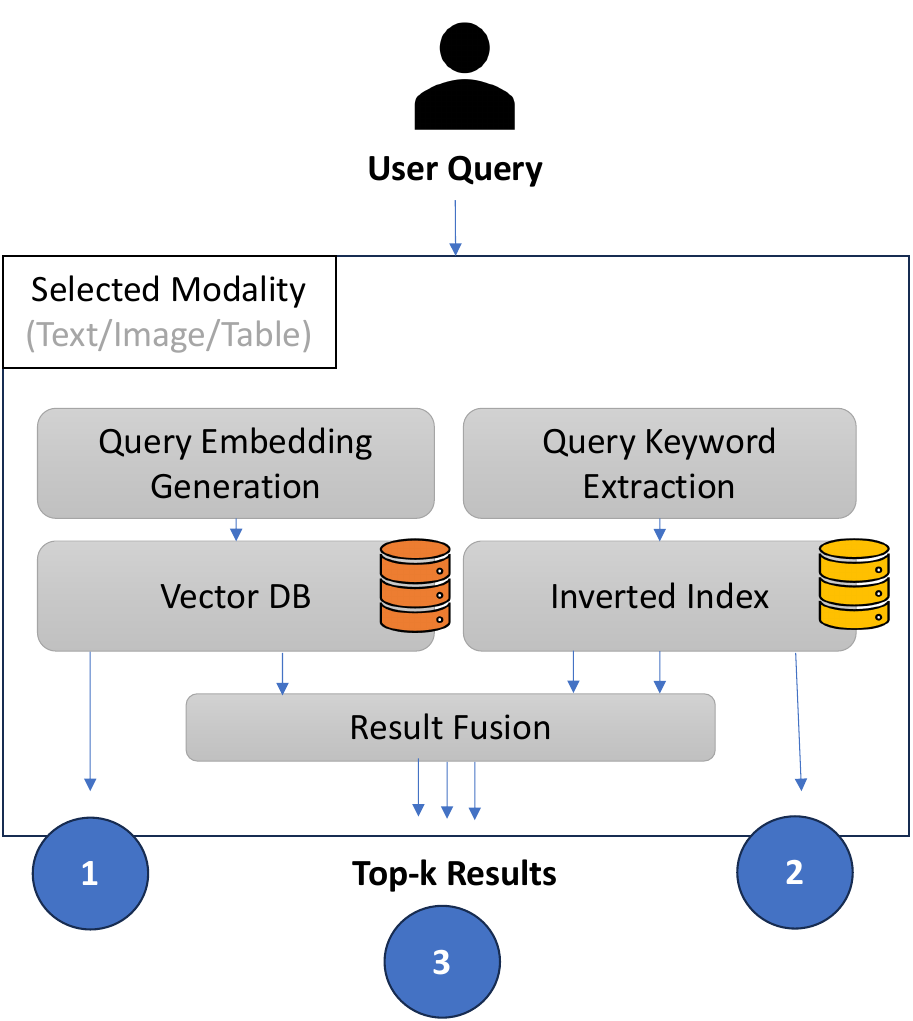}
    \caption{Search Pipelines: 1. The first pipeline takes in the user's query and modality and searches the appropriate vector DB, 2. The second pipeline does a keyword search using the BM25 scoring model, 3. The third pipeline is a hybrid method that combines and reranks results from the previous two searches.}
    \label{fig:search-arch}
\end{figure}


The search takes a user query in natural language and can return results in three different modalities: text, image, and table. We have analyzed three different retrieval systems, a keyword-based search,  an embedding-based search and a hybrid approach. Figure \ref{fig:search-arch} describes these pipelines.
 

\subsubsection*{Keyword-Based Search}
The query is analyzed to extract keywords, which are then used to search the corpus. 
The corpus, consisting of text extracted by the content extraction pipeline (Section \ref{sec:pipeline}) is indexed at the page level for each document, using Apache Solr \cite{solr}. The extracted images and tables are indexed using their extracted textual context.

The most important keywords from the user's query are extracted using standard methods. For each query, the top-$k$ results, scored using BM25 \cite{bm25Solr} \cite{robertson2009probabilistic} are returned. We expect this method to be beneficial for queries with archaeology specific words.


\subsubsection*{Embedding-Based Search}
This is a multi-modal embeddings based search system. Embeddings are generated using modality-specific models.
For textual data, embeddings are generated using pre-trained sentence transformer all-MiniLM-L6-v2 \cite{sentenceTransformerMiniLMPaper,sentenceTranformerSBERT}; for image data and its extracted textual context, embeddings for both images and their associated textual context are generated using CLIP \cite{clip}; for tabular data and its associated context, embeddings are generated using TAPAS \cite{herzig2020tapas}. These embeddings are stored in appropriated indexes - (Solr \cite{solr}, Postgres+pgvector \cite{pgvector}) along with associated metadata (eg. document name, path, etc). 

User query in natural language is encoded using a modality-specific model. Top-$k$ results corresponding to most similar embeddings associated with the selected modality in the vector database are scored using cosine similarity and returned. \footnote{Note that the users have to choose the modality of the result. We are working on methods to eliminate the need}



\subsubsection*{Hybrid search}
The results of these retrieval systems are ranked using their scores and are merged using reciprocal rank fusion \cite{rrf}. Top-$k$ of the merged results based on the rank are returned.

\subsubsection*{Examples} 
\begin{figure*} [ht!]
    \begin{subfigure}{0.33\textwidth}
        \centering
        \includegraphics[width=\linewidth]{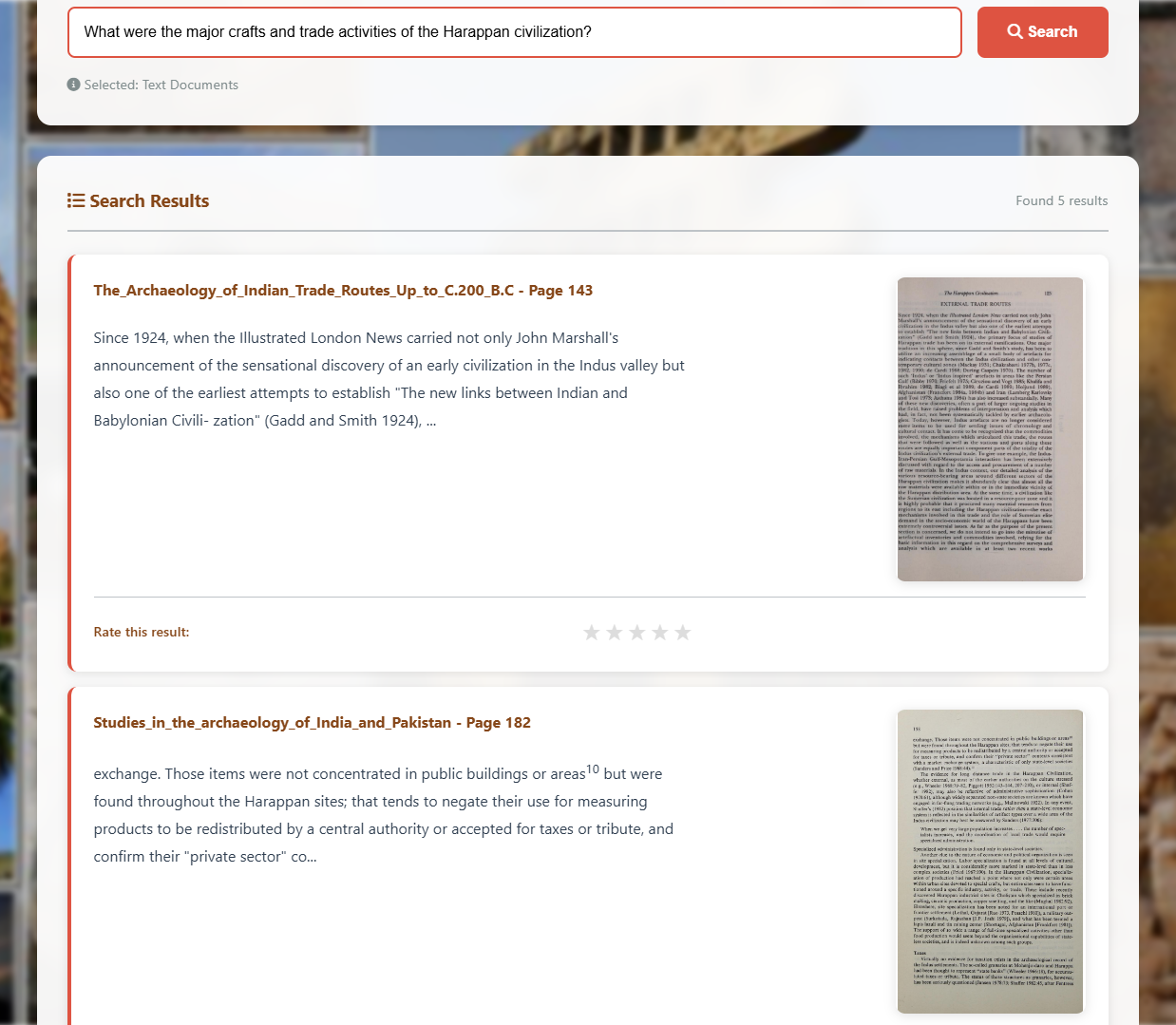}
        \caption{Text modality (Query: "Major trade activities of the Harappan civilization")}
        \label{fig:text-result}
    \end{subfigure}
    \begin{subfigure}{0.33\textwidth}
        \centering
        \includegraphics[width=\linewidth]{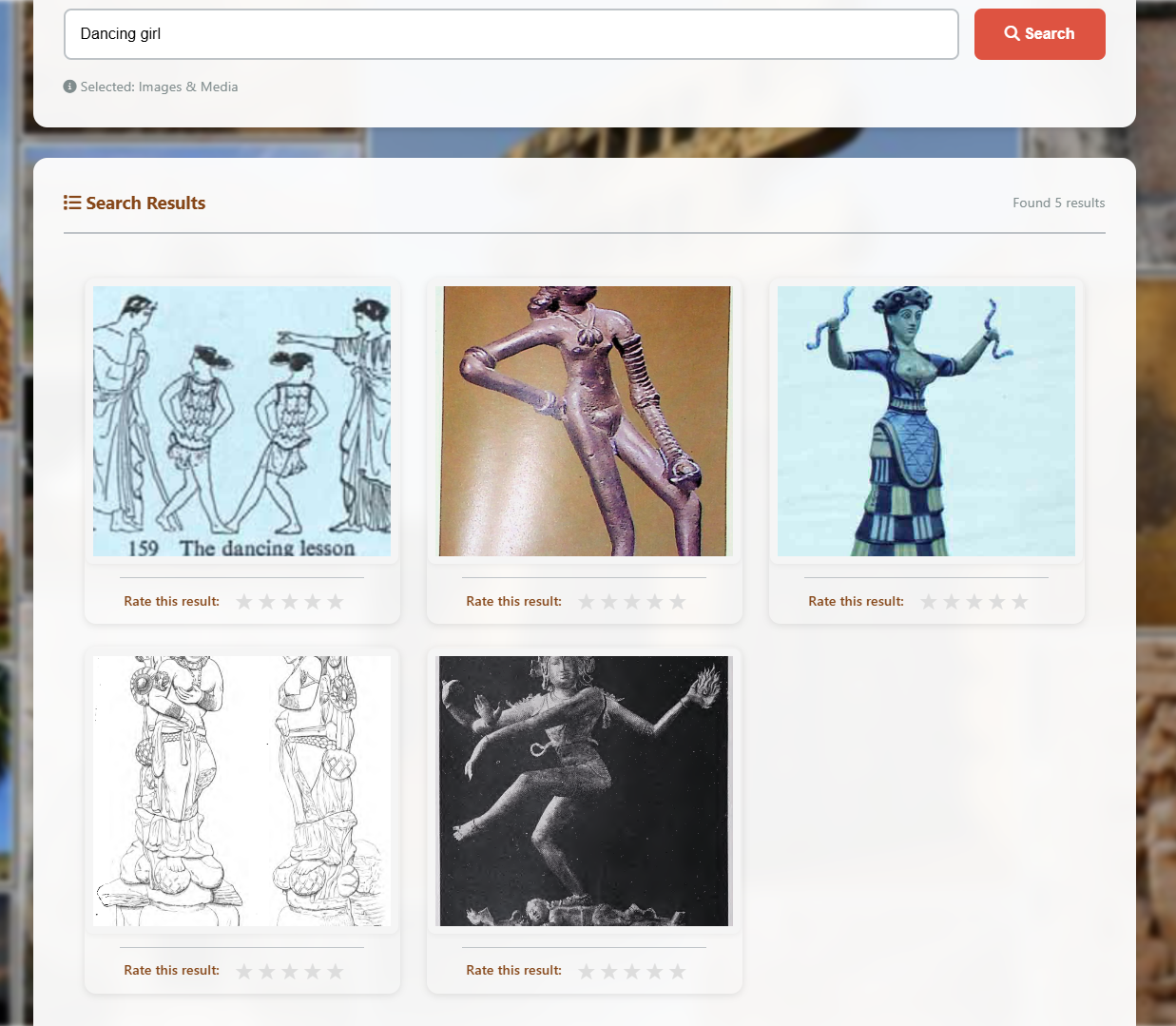}
        \caption{Image modality \\(Query: "Dancing Girl")}
        \label{fig:image-result}
    \end{subfigure}
    \begin{subfigure}{0.33\textwidth}
        \centering
        \includegraphics[width=\linewidth]{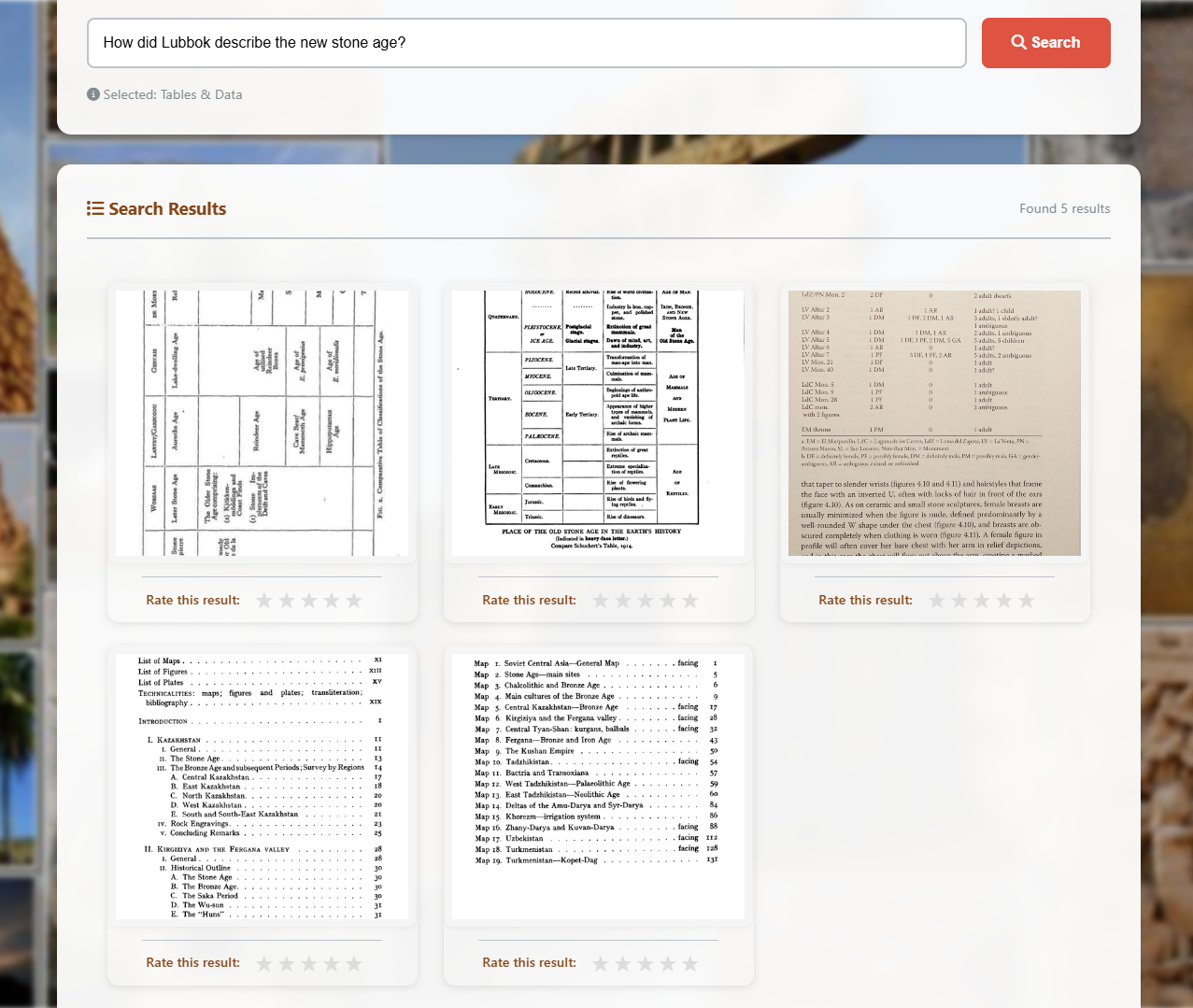}
        \caption{Table modality (Query: "How did Lubbock describe the New Stone Age")}
        \label{fig:table-result}
    \end{subfigure}
\caption{Interface of our search system: showing results for queries in different modalities}
\label{fig:search-interface}
    
\end{figure*}

Figure 5 shows examples of the result by hybrid search for each of the three modality types. For text, the query "Major trade
activities of the Harappan civilization" in Figure \ref{fig:text-result} returns snippets from a book on the Archaeology of Indian Trade Routes as the first result.
For image modality, the query "photo: dancing girl" in Figure
\ref{fig:image-result} shows at least 3 photos of dancing girls, including the famous dancing girl from Mohenjodaro (second result).
For table modality, the query “How did Lubbock describe the New Stone Age?” returns a comparative table of classifications of the stone age, showing how different scholars, including Lubbock, classified prehistoric
periods (the first result in Figure \ref{fig:table-result}).

\section{Evaluation}
\label{sec:eval}

\begin{table*}[!htbp]
    \vspace*{0.6cm}
    \centering
    \begin{tabular}{|p{6cm}|c|c|c|c|c|}
    \hline
       {\bf Search Type}  & {\bf P@5} & {\bf P@3} & {\bf P@1} & {\bf MRR} \\
       \hline
        Keyword-Based Search & 0.238 & 0.301 & 0.429 & 0.508 \\
        \hline
        Embedding-Based Search & 0.438 & 0.476 & 0.619 & 0.754 \\
        \hline
         Hybrid search  & 0.346 & 0.41 & 0.538 & 0.726 \\
    \hline
    \end{tabular}
    \vspace{0.1cm}
    \caption{Benchmark Evaluation results}
    \label{tab:metric}
\end{table*}

We are not aware of any standard benchmarks for archaeological search. Therefore, we created our own benchmarks and report on our preliminary results on these benchmarks.

\subsubsection*{Benchmark}

We solicited the help of a professional archaeologist in creating and refining 15 text 11 image queries and 4 table queries that would be of interest to archaeologists. 





\subsubsection*{Results and Analysis} We generated the top-5 results and asked PhD students in our group to evaluate the relevance of each result.
We calculated P@5, P@3, P@1 and MRR for each of the search pipelines: Keyword-based, embedding-based and hybrid search. The results are tabulated in Table \ref{tab:metric}. Our results show that the embedding-based search results are better overall. The MRR and P@1 numbers indicate that relevant results are ranked relatively high. However, in our preliminary analysis, we observed that for queries with archaeology-specific terms (example: "Harappa", "New Stone Age", etc.), the keyword-based search pipeline gave the best results.


\section{Conclusion and Future Works}
\label{sec:conc}
In this paper, we described SARCH, our multi-modal search system for archaeological archive data. The data consists of pdf scans of various documents and is of varying quality. We developed a content extraction system to extract text, images, and tables from these scans. To query this data, we implemented three search pipelines consisting of both embedding-based search as well as keyword search. Our preliminary results are promising.

Our future work will focus on both improved content and context extraction (especially from maps and tables) as well as improved search pipelines that use instruction-tuned models that eliminate the need for the user to select a modality. We plan to further investigate which queries perform well with just keyword context and which need image representations as well.

\section{GenAI Usage Disclosure}
We have used GenAI tools to help write code for the system described here. We have also used GenAI tools to correct typos and grammar in writing this paper.

\bibliographystyle{plain}

\begin{thebibliography}{10}

\bibitem{solr}
Apache solr.
\newblock https://solr.apache.org/.

\bibitem{bm25Solr}
Apache solr reference guide.
\newblock https://solr.apache.org/guide/solr/latest/indexing-guide/schema-elements.html\#similarity.

\bibitem{arachne}
Arachne.
\newblock https://arachne.dainst.org.

\bibitem{ads}
Archaeology data service.
\newblock https://archaeologydataservice.ac.uk.

\bibitem{tdar}
The digital archaeological record.
\newblock https://www.tdar.org.

\bibitem{sentenceTranformerSBERT}
Pretrained models - sentence transformer documentation.
\newblock \url{https://www.sbert.net/docs/sentence_transformer/pretrained_models.html}.
\newblock Last Accessed: 2025-06-16.

\bibitem{surya}
Vik paruchuri. 2024b. surya: Accurate line-by-line text detection and recognition in complex documents.

\bibitem{beautifulSoup}
Beautiful soup.
\newblock \url{https://beautiful-soup-4.readthedocs.io/en/latest/}, 2015.
\newblock Last Accessed: 2025-06-17.

\bibitem{pgvector}
pgvector: Open-source vector similarity search for postgres.
\newblock \url{https://github.com/pgvector/pgvector}, 2021.
\newblock Last Accessed: 2025-06-16.

\bibitem{textblob}
Textblob.
\newblock \url{https://textblob.readthedocs.io/en/dev/quickstart.html}, 2025.
\newblock Last Accessed: 2025-06-17.

\bibitem{bai2025qwen2}
Shuai Bai, Keqin Chen, Xuejing Liu, Jialin Wang, Wenbin Ge, Sibo Song, Kai Dang, Peng Wang, Shijie Wang, Jun Tang, et~al.
\newblock Qwen2. 5-vl technical report.
\newblock {\em arXiv preprint arXiv:2502.13923}, 2025.

\bibitem{bradski2000opencv}
Gary Bradski.
\newblock The opencv library.
\newblock {\em Dr. Dobb's Journal: Software Tools for the Professional Programmer}, 25(11):120--123, 2000.

\bibitem{rrf}
Gordon~V. Cormack, Charles L.~A. Clarke, and Stefan B{\"{u}}ttcher.
\newblock Reciprocal rank fusion outperforms condorcet and individual rank learning methods.
\newblock In James Allan, Javed~A. Aslam, Mark Sanderson, ChengXiang Zhai, and Justin Zobel, editors, {\em Proceedings of the 32nd Annual International {ACM} {SIGIR} Conference on Research and Development in Information Retrieval, {SIGIR} 2009, Boston, MA, USA, July 19-23, 2009}, pages 758--759. {ACM}, 2009.

\bibitem{herzig2020tapas}
Jonathan Herzig, Pawe{\l}~Krzysztof Nowak, Thomas M{\"u}ller, Francesco Piccinno, and Julian~Martin Eisenschlos.
\newblock Tapas: Weakly supervised table parsing via pre-training.
\newblock {\em arXiv preprint arXiv:2004.02349}, 2020.

\bibitem{otsu1979}
Nobuyuki Otsu.
\newblock A threshold selection method from gray-level histograms.
\newblock {\em IEEE Transactions on Systems, Man, and Cybernetics}, 9(1):62--66, 1979.

\bibitem{suryaocr}
Vikas Paruchuri and Datalab Team.
\newblock Surya: A lightweight document ocr and analysis toolkit.
\newblock \url{https://github.com/VikParuchuri/surya}, 2025.
\newblock GitHub repository.

\bibitem{clip}
Alec Radford, Jong~Wook Kim, Chris Hallacy, Aditya Ramesh, Gabriel Goh, Sandhini Agarwal, Girish Sastry, Amanda Askell, Pamela Mishkin, Jack Clark, Gretchen Krueger, and Ilya Sutskever.
\newblock Learning transferable visual models from natural language supervision.
\newblock In Marina Meila and Tong Zhang, editors, {\em Proceedings of the 38th International Conference on Machine Learning, {ICML} 2021, 18-24 July 2021, Virtual Event}, volume 139 of {\em Proceedings of Machine Learning Research}, pages 8748--8763. {PMLR}, 2021.

\bibitem{robertson2009probabilistic}
Stephen Robertson, Hugo Zaragoza, et~al.
\newblock The probabilistic relevance framework: Bm25 and beyond.
\newblock {\em Foundations and Trends{\textregistered} in Information Retrieval}, 3(4):333--389, 2009.

\bibitem{sentenceTransformerMiniLMPaper}
Wenhui Wang, Furu Wei, Li~Dong, Hangbo Bao, Nan Yang, and Ming Zhou.
\newblock Minilm: Deep self-attention distillation for task-agnostic compression of pre-trained transformers.
\newblock {\em Advances in neural information processing systems}, 33:5776--5788, 2020.

\end{thebibliography}

\clearpage
\onecolumn        
\begin{center}
\Large \textbf{Appendix}
\end{center}
\vspace{0.8em}
\addcontentsline{toc}{section}{Appendix}
\vspace{0.5em}

\begin{center}
\renewcommand{\arraystretch}{1.25}
\setlength{\tabcolsep}{10pt}
\footnotesize

\begin{tabular}{|p{3cm}|p{13.5cm}|}
    \hline
    \textbf{Modality} & \textbf{Queries} \\ \hline

    \textbf{Image} &
    1. Dancing girl image in Mohenjo-daro \newline 
    2. Major monuments from Harappan civilization \newline 
    3. What kind of vessels did Indus people use for their food \newline 
    4. Show the map for Rigvedic era of Harappan civilization \newline 
    5. Workmen's quarters in Harappan civilization \newline 
    6. Urban planning of Harappan civilization \newline
    7. Map for sites in Gujarat as part of Harappan tradition \newline 
    8. Top sites of excavation at Lothal  \newline 
    9. Bricks used for building the houses in Harappan civilization \newline 
    10. Major artworks from Harappan civilization \newline 
    11. Top sites of excavation at Kalibangan \\ \hline 

    \textbf{Text} &
    1. Primary crops of the Harappan civilization \newline 
    2. What were the religious beliefs of Harappan people \newline 
    3. Where were the workmen's quarters discovered in Harappa \newline 
    4. Major crafts and trade in Harappan civilization \newline 
    5. What are the important features of Harappan culture  \newline
    6. What are the ornaments and jewellery in Harappa \newline 
    7. Chief source of copper for Harappan people \newline 
    8. Urban planning of Harappan civilization \newline 
    9. Scriptures from Harappan civilization \newline 
    10. Animals in Harappan civilization \newline 
    11. Major artworks from Harappan civilization \newline
    12. What kind of vessels did Indus people use for their food \newline 
    13. Major agricultural crops for Harappan civilization,  Primary crops of the Harappan civilization \newline 
    14. Give the evidence of fire worships in Harappan civilization \newline
    15. Weapons used in Harappan civilization \\ \hline 

    \textbf{Table} &
    1. Description of Lubbock in the New Stone Age? \newline
    2. Which ancient settlements have produced steatite bead artifacts in excavations? \newline
    3. How are nails and knives distributed in Sub-period IIB vs. IIA at Bharadvaja Ashrama? \newline
    4. What is the dominant life in the Holocene epoch? \\ \hline
   
\end{tabular}
\captionof{table}{Modality-wise Benchmark Queries used for Evaluation}

\label{tab:appendix-queries}
\end{center}

\twocolumn          

\end{document}